\pdfoutput=1

\documentclass[12pt,a4paper]{article}

\usepackage{ifthen} 
\newboolean{pdflatex}
\setboolean{pdflatex}{true} 

\newboolean{articletitles}
\setboolean{articletitles}{true} 

\newboolean{uprightparticles}
\setboolean{uprightparticles}{false} 

\newcommand{\Acp}{\ensuremath{\mathcal{A}_{\CP}}\xspace}
\newcommand{\ARAW}{\ensuremath{A}\xspace}

\newcommand{\DorDsp}{\ensuremath{\D_{(\squark)}^+}\xspace}
\newcommand{\DorDsm}{\ensuremath{\D_{(\squark)}^-}\xspace}
\newcommand{\DsKSpi}{\ensuremath{\Dsp\to\KS\pip}\xspace}
\newcommand{\DKSpi}{\ensuremath{\Dp\to\KS\pip}\xspace}
\newcommand{\DsKSK}{\ensuremath{\Dsp\to\KS\Kp}\xspace}
\newcommand{\DKSK}{\ensuremath{\Dp\to\KS\Kp}\xspace}
\newcommand{\Dsphipi}{\ensuremath{\Dsp\to\phi\pip}\xspace}
\newcommand{\Dphipi}{\ensuremath{\Dp\to\phi\pip}\xspace}
\newcommand{\DorDspKSpi}{\ensuremath{\DorDsp\to\KS\pip}\xspace}
\newcommand{\DorDspKSK}{\ensuremath{\DorDsp\to\KS\Kp}\xspace}
\newcommand{\DorDspphipi}{\ensuremath{\DorDsp\to\phi\pip}\xspace}
\newcommand{\tip}{\ensuremath{\text{TIP}}\xspace}

\newcommand{\AcpDsKSpiRes}{1.3}
\newcommand{\AcpDsKSpiStat}{1.9}
\newcommand{\AcpDsKSpiSyst}{0.5}
\newcommand{\AcpDsKSpiUnit}{10^{-3}}

\newcommand{\AcpDKSKRes}{-0.09}
\newcommand{\AcpDKSKStat}{0.65}
\newcommand{\AcpDKSKSyst}{0.48}
\newcommand{\AcpDKSKUnit}{10^{-3}}

\newcommand{\AcpDphipiRes}{0.05}
\newcommand{\AcpDphipiStat}{0.42}
\newcommand{\AcpDphipiSyst}{0.29}
\newcommand{\AcpDphipiUnit}{10^{-3}}

\newcommand{\mypaperversion}{}
\newcommand{\mydate}{March 5, 2019}
\newcommand{\mylhcbpapernumber}{LHCb-PAPER-2019-002}
\newcommand{\mycernpapernumber}{CERN-EP-2019-027}

\def\paperauthors{LHCb collaboration} 
\def\paperasciititle{Search for CP violation in Ds->KSpi, D->KSK and D->phipi decays} 
\def\papertitle{Search for \CP violation in \DsKSpi, \DKSK and \Dphipi decays} 
\def\paperkeywords{{High Energy Physics}, {LHCb}} 
\def\papercopyright{\the\year\ CERN for the benefit of the LHCb collaboration} 
\def\paperlicence{CC-BY-4.0 licence}
\def\paperlicenceurl{https://creativecommons.org/licenses/by/4.0/}

\usepackage[top=1in, bottom=1.25in, left=1in, right=1in]{geometry}

\usepackage{microtype}
\usepackage{lineno}  
\usepackage{xspace} 
\usepackage{caption} 

\usepackage{graphicx}  
\usepackage{color}
\usepackage{colortbl}
\graphicspath{{./figs/}{./figs/paper/}{./figs/supplementary/}} 
\DeclareGraphicsExtensions{.pdf,.PDF,png,.PNG}

\usepackage{amsmath} 
\usepackage{amssymb}
\usepackage{amsfonts}
\usepackage{upgreek} 
\usepackage{overpic}
\newcommand*\patchAmsMathEnvironmentForLineno[1]{%
\expandafter\let\csname old#1\expandafter\endcsname\csname #1\endcsname
\expandafter\let\csname oldend#1\expandafter\endcsname\csname
end#1\endcsname
 \renewenvironment{#1}%
   {\linenomath\csname old#1\endcsname}%
   {\csname oldend#1\endcsname\endlinenomath}%
}
\newcommand*\patchBothAmsMathEnvironmentsForLineno[1]{%
  \patchAmsMathEnvironmentForLineno{#1}%
  \patchAmsMathEnvironmentForLineno{#1*}%
}
\AtBeginDocument{%
\patchBothAmsMathEnvironmentsForLineno{equation}%
\patchBothAmsMathEnvironmentsForLineno{align}%
\patchBothAmsMathEnvironmentsForLineno{flalign}%
\patchBothAmsMathEnvironmentsForLineno{alignat}%
\patchBothAmsMathEnvironmentsForLineno{gather}%
\patchBothAmsMathEnvironmentsForLineno{multline}%
\patchBothAmsMathEnvironmentsForLineno{eqnarray}%
}

\usepackage{hyperxmp}

\usepackage[pdftex,
            pdfauthor={\paperauthors},
            pdftitle={\paperasciititle},
            pdfkeywords={\paperkeywords},
            pdfcopyright={Copyright (C) \papercopyright},
            pdflicenseurl={\paperlicenceurl}]{hyperref}

\usepackage[all]{hypcap} 

\usepackage{xspace} 
\usepackage{upgreek}

\newcommand{\offsetoverline}[2][0.1em]{\kern #1\overline{\kern -#1 #2}}%

\def\MagUp {\mbox{\em Mag\kern -0.05em Up}\xspace}

\ifthenelse{\boolean{uprightparticles}}%
{

 \def\Ppi         {\ensuremath{\uppi}\xspace}

 \def\PDelta      {\ensuremath{\Delta}\xspace}                 
 \def\PXi         {\ensuremath{\Xi}\xspace}                 
 \def\PLambda     {\ensuremath{\Lambda}\xspace}                 
 \def\PSigma      {\ensuremath{\Sigma}\xspace}                 
 \def\POmega      {\ensuremath{\Omega}\xspace}                 
 \def\PUpsilon    {\ensuremath{\Upsilon}\xspace}

 \def\PB      {\ensuremath{\mathrm{B}}\xspace}                 
                  
 \def\PD      {\ensuremath{\mathrm{D}}\xspace}

 \def\PK      {\ensuremath{\mathrm{K}}\xspace}

 \def\Pb      {\ensuremath{\mathrm{b}}\xspace}                 
 \def\Pc      {\ensuremath{\mathrm{c}}\xspace}                 
 \def\Pd      {\ensuremath{\mathrm{d}}\xspace}

 \def\Pi      {\ensuremath{\mathrm{i}}\xspace}

 \def\Pp      {\ensuremath{\mathrm{p}}\xspace}

 \def\Ps      {\ensuremath{\mathrm{s}}\xspace}                 
                  
 \def\Pu      {\ensuremath{\mathrm{u}}\xspace}

}
{

 \def\Ppi         {\ensuremath{\pi}\xspace}

 \mathchardef\PDelta="7101
 \mathchardef\PXi="7104
 \mathchardef\PLambda="7103
 \mathchardef\PSigma="7106
 \mathchardef\POmega="710A
 \mathchardef\PUpsilon="7107
                  
 \def\PB      {\ensuremath{B}\xspace}                 
                  
 \def\PD      {\ensuremath{D}\xspace}

 \def\PK      {\ensuremath{K}\xspace}

 \def\Pb      {\ensuremath{b}\xspace}                 
 \def\Pc      {\ensuremath{c}\xspace}                 
 \def\Pd      {\ensuremath{d}\xspace}

 \def\Pi      {\ensuremath{i}\xspace}

 \def\Pp      {\ensuremath{p}\xspace}

 \def\Ps      {\ensuremath{s}\xspace}                 
                  
 \def\Pu      {\ensuremath{u}\xspace}

}

\makeatletter
\ifcase \@ptsize \relax
  \newcommand{\miniscule}{\@setfontsize\miniscule{4}{5}}
\or
  \newcommand{\miniscule}{\@setfontsize\miniscule{5}{6}}
\or
  \newcommand{\miniscule}{\@setfontsize\miniscule{5}{6}}
\fi
\makeatother

\DeclareRobustCommand{\optbar}[1]{\shortstack{{\miniscule (\rule[.5ex]{1.25em}{.18mm})}
  \\ [-.7ex] $#1$}}

\def\uquark    {{\ensuremath{\Pu}}\xspace}

\def\dquark    {{\ensuremath{\Pd}}\xspace}
\def\dquarkbar {{\ensuremath{\overline \dquark}}\xspace}

\def\squark    {{\ensuremath{\Ps}}\xspace}
\def\squarkbar {{\ensuremath{\overline \squark}}\xspace}

\def\cquark    {{\ensuremath{\Pc}}\xspace}

\def\bquark    {{\ensuremath{\Pb}}\xspace}

\def\pion   {{\ensuremath{\Ppi}}\xspace}

\def\pip    {{\ensuremath{\pion^+}}\xspace}
\def\pim    {{\ensuremath{\pion^-}}\xspace}

\def\kaon    {{\ensuremath{\PK}}\xspace}
  \def\Kbar    {{\kern 0.2em\overline{\kern -0.2em \PK}{}}\xspace}

\def\KorKbar {\kern 0.18em\optbar{\kern -0.18em K}{}\xspace}
\def\Kz      {{\ensuremath{\kaon^0}}\xspace}
\def\Kzb     {{\ensuremath{\Kbar{}^0}}\xspace}
\def\Kp      {{\ensuremath{\kaon^+}}\xspace}
\def\Km      {{\ensuremath{\kaon^-}}\xspace}

\def\KS      {{\ensuremath{\kaon^0_{\mathrm{S}}}}\xspace}

  \def\Dbar    {{\kern 0.2em\overline{\kern -0.2em \PD}{}}\xspace}
\def\D       {{\ensuremath{\PD}}\xspace}

\def\DorDbar {\kern 0.18em\optbar{\kern -0.18em D}{}\xspace}

\def\Dp      {{\ensuremath{\D^+}}\xspace}

\def\Dsp     {{\ensuremath{\D^+_\squark}}\xspace}

\def\Bbar    {{\ensuremath{\kern 0.18em\overline{\kern -0.18em \PB}{}}}\xspace}

\def\BorBbar    {\kern 0.18em\optbar{\kern -0.18em B}{}\xspace}

\def\Y#1S{\ensuremath{\PUpsilon{(#1S)}}\xspace}

\def\proton      {{\ensuremath{\Pp}}\xspace}

\def\Lz          {{\ensuremath{\PLambda}}\xspace}

\def\LorLbar     {\kern 0.18em\optbar{\kern -0.18em \PLambda}{}\xspace}

\def\Lc          {{\ensuremath{\Lz^+_\cquark}}\xspace}

\def\to                 {\ensuremath{\rightarrow}\xspace}

\def\order   {{\ensuremath{\mathcal{O}}}\xspace}

\def\CP                {{\ensuremath{C\!P}}\xspace}

\def\AT#1     {\ensuremath{A_{\mathrm{T}}^{#1}}\xspace}           

\def\C#1      {\ensuremath{\mathcal{C}_{#1}}\xspace}                       
\def\Cp#1     {\ensuremath{\mathcal{C}_{#1}^{'}}\xspace}                    
\def\Ceff#1   {\ensuremath{\mathcal{C}_{#1}^{\mathrm{(eff)}}}\xspace}        
\def\Cpeff#1  {\ensuremath{\mathcal{C}_{#1}^{'\mathrm{(eff)}}}\xspace}       
\def\Ope#1    {\ensuremath{\mathcal{O}_{#1}}\xspace}                       
\def\Opep#1   {\ensuremath{\mathcal{O}_{#1}^{'}}\xspace}                    

\newcommand{\nospaceunit}[1]{\ensuremath{\text{#1}}}       
\newcommand{\aunit}[1]{\ensuremath{\text{\,#1}}}       

\newcommand{\tev}{\aunit{Te\kern -0.1em V}\xspace}
\newcommand{\gev}{\aunit{Ge\kern -0.1em V}\xspace}
\newcommand{\mev}{\aunit{Me\kern -0.1em V}\xspace}
\newcommand{\kev}{\aunit{ke\kern -0.1em V}\xspace}
\newcommand{\ev}{\aunit{e\kern -0.1em V}\xspace}
\newcommand{\mevc}{\ensuremath{\aunit{Me\kern -0.1em V\!/}c}\xspace}
\newcommand{\gevc}{\ensuremath{\aunit{Ge\kern -0.1em V\!/}c}\xspace}
\newcommand{\mevcc}{\ensuremath{\aunit{Me\kern -0.1em V\!/}c^2}\xspace}
\newcommand{\gevcc}{\ensuremath{\aunit{Ge\kern -0.1em V\!/}c^2}\xspace}

\def\mum  {\ensuremath{\,\upmu\nospaceunit{m}}\xspace}

\def\fb   {\ensuremath{\aunit{fb}}\xspace}
\def\invfb   {\ensuremath{\fb^{-1}}\xspace}

\def\order{{\ensuremath{\mathcal{O}}}\xspace}
\newcommand{\chisq}{\ensuremath{\chi^2}\xspace}

\def\gsim{{~\raise.15em\hbox{$>$}\kern-.85em
          \lower.35em\hbox{$\sim$}~}\xspace}
\def\lsim{{~\raise.15em\hbox{$<$}\kern-.85em
          \lower.35em\hbox{$\sim$}~}\xspace}

\def\sPlot{\mbox{\em sPlot}\xspace}

\xspace

\def\tell1  {TELL1\xspace}
\def\ukl1   {UKL1\xspace}

\newcommand{\ie}{\mbox{\itshape i.e.}\xspace}

\usepackage{cite} 
\usepackage{mciteplus}

\usepackage[capitalise]{cleveref}
\Crefname{figure}{Figure}{Figures}

\usepackage{booktabs}

\begin{document}
\renewcommand{\thefootnote}{\fnsymbol{footnote}}
\setcounter{footnote}{1}

\begin{titlepage}
\pagenumbering{roman}

\vspace*{-1.5cm}
\centerline{\large EUROPEAN ORGANIZATION FOR NUCLEAR RESEARCH (CERN)}
\vspace*{1.5cm}
\noindent
\begin{tabular*}{\linewidth}{lc@{\extracolsep{\fill}}r@{\extracolsep{0pt}}}
\vspace*{-1.5cm}\mbox{\!\!\!\includegraphics[width=.14\textwidth]{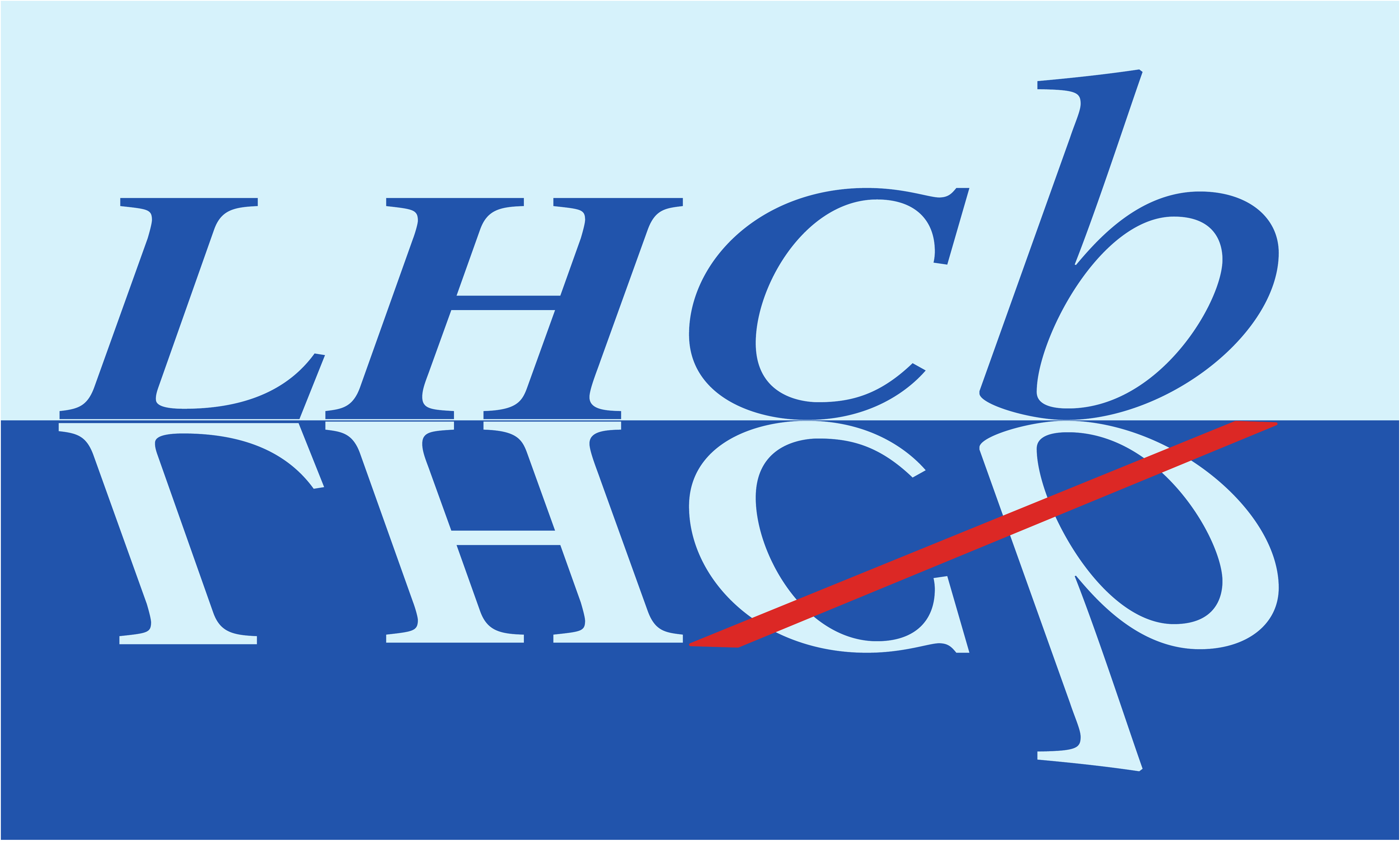}} & & \\
 & & \mycernpapernumber \\
 & & \mylhcbpapernumber \\
 & & \mydate \\ 
 & & \mypaperversion\\
\end{tabular*}

\vspace*{3.0cm}

{\normalfont\bfseries\boldmath\huge
\begin{center}
  \papertitle 
\end{center}
}

\vspace*{1.3cm}

\begin{center}
\paperauthors\footnote{Authors are listed at the end of this paper.}
\end{center}

\vspace{\fill}

\begin{abstract}
\noindent A search for charge-parity (\CP) violation in Cabibbo-suppressed \DsKSpi, \DKSK and \Dphipi decays is reported using proton-proton collision data, corresponding to an integrated luminosity of 3.8\invfb, collected at a center-of-mass energy of 13\tev with the LHCb detector. High-yield samples of kinematically and topologically similar Cabibbo-favored \DorDsp decays are analyzed to subtract nuisance asymmetries due to production and detection effects, including those induced by \CP violation in the neutral kaon system. The results are
\begin{align*}
\Acp(\DsKSpi) &=\left(\phantom{-}\AcpDsKSpiRes\phantom{0}\pm\AcpDsKSpiStat\phantom{0}\pm\AcpDsKSpiSyst\phantom{0}\right)\times\AcpDsKSpiUnit,\\
\Acp(\DKSK) &=\left(\AcpDKSKRes\pm\AcpDKSKStat\pm\AcpDKSKSyst\right)\times\AcpDKSKUnit,\\
\Acp(\Dphipi) &=\left(\phantom{-}\AcpDphipiRes\pm\AcpDphipiStat\pm\AcpDphipiSyst\right)\times\AcpDphipiUnit,
\end{align*}
where the first uncertainties are statistical and the second systematic. They are the most precise measurements of these quantities to date, and are consistent with \CP symmetry. A combination with previous LHCb measurements, based on data collected at 7 and 8\tev, is also reported.

\end{abstract}

\vspace*{1.5cm}

\begin{center}
Published in \href{https://journals.aps.org/prl/abstract/10.1103/PhysRevLett.122.191803}{Phys.\ Rev.\ Lett.\ {\bf 122} (2019) 191803}
\end{center}

\vspace{\fill}

{\footnotesize 
\centerline{\copyright~\papercopyright. \href{\paperlicenceurl}{\paperlicence}.}}
\vspace*{2mm}

\end{titlepage}


\newpage
\setcounter{page}{2}
\mbox{~}

\cleardoublepage

\renewcommand{\thefootnote}{\arabic{footnote}}
\setcounter{footnote}{0}


\pagestyle{plain} 
\setcounter{page}{1}
\pagenumbering{arabic}


\noindent 
Violation of charge-parity (\CP) symmetry arises in the Standard Model (SM) of particle physics through the complex phase of the Cabibbo--Kobayashi--Maskawa (CKM) quark-mixing matrix~\cite{Cabibbo:1963yz,Kobayashi:1973fv}. \CP violation is well established in $K$- and $B$-meson systems~\cite{Christenson:1964fg,Aubert:2004qm,Chao:2004mn,LHCb-PAPER-2013-018,LHCb-PAPER-2012-001}, and has been observed only recently in charm decays~\cite{LHCb-PAPER-2019-006}. \CP violation in charm decays can arise from the interference between tree- and loop-level diagrams through Cabibbo-suppressed $\cquark\to\dquark\dquarkbar\uquark$ and $\cquark\to\squark\squarkbar\uquark$ transition amplitudes. In the loop-level processes, contributions from physics beyond the SM may arise that can lead to additional sources of \CP violation~\cite{Grossman:2006jg}. However, the expected SM contribution is difficult to compute due to the presence of low-energy strong-interaction effects, with current predictions spanning several orders of magnitude~\cite{Golden:1989qx,Buccella:1994nf,Bianco:2003vb,Grossman:2006jg,Artuso:2008vf}. A promising handle to determine the origin of possible \CP-violation signals are correlations between \CP asymmetries in flavor-$SU(3)$ related decays~\cite{Pirtskhalava:2011va,Cheng:2012wr,Feldmann:2012js,Li:2012cfa,Franco:2012ck,Brod:2012ud,Atwood:2012ac,Hiller:2012xm,Muller:2015rna}. Particularly interesting in this respect are \Dsp and \Dp decays to two-body (or quasi two-body) final states, such as \DsKSpi, \DKSK and \Dphipi.\footnote{The inclusion of charge-conjugate processes is implied throughout this Letter, unless stated otherwise.} Searches for \CP violation in these modes have been performed by the CLEO~\cite{Mendez:2009aa}, BaBar~\cite{Lees:2012jv,Lees:2012nn}, Belle~\cite{Ko:2010ng,Ko:2012uh,Staric:2011en} and LHCb~\cite{LHCb-PAPER-2012-052,LHCb-PAPER-2014-018} collaborations. No evidence for \CP violation has been found within a precision of a few per mille.

This Letter presents measurements of \CP asymmetries in \DsKSpi, \DKSK and \Dphipi decays performed using proton-proton collision data collected with the LHCb detector between 2015 and 2017 at a center-of-mass energy of 13\tev, and corresponding to an integrated luminosity of $3.8\invfb$. In the presence of a \KS meson in the final state, a \CP asymmetry is expected to be induced by \Kz--\Kzb mixing~\cite{Lipkin:1999qz}. This effect is well known and predictable, allowing for a precise measurement of \CP violation in the charm-quark transition. The \Dphipi decay is reconstructed with the $\phi\to\Kp\Km$ mode. Several intermediate states contribute to the $\Dp\to\Kp\Km\pip$ decay amplitude~\cite{PDG2018}. In this Letter, no attempt is made to separate them through an amplitude analysis and the measurement is performed by simply restricting the $\Kp\Km$ pair to the mass region around the $\phi(1020)$ resonance.

The \CP asymmetry of a \DorDsp meson decaying to the final state $f^+$ is defined as
\begin{equation}
\Acp(\DorDsp\to f^+) \equiv \frac{\Gamma(\DorDsp\to f^+)-\Gamma(\DorDsm\to f^-)}{\Gamma(\DorDsp\to f^+)+\Gamma(\DorDsm\to f^-)},
\end{equation}
where $\Gamma$ is the partial decay rate. If \CP symmetry is violated in the decay, $\Acp\neq0$. An experimentally convenient quantity to measure is the ``raw'' asymmetry of the observed yields $N$,
\begin{equation}
\ARAW(\DorDsp\to f^+) \equiv \frac{N(\DorDsp\to f^+)-N(\DorDsm\to f^-)}{N(\DorDsp\to f^+)+N(\DorDsm\to f^-)}.
\end{equation}
The raw asymmetry can be approximated as
\begin{equation}
\ARAW(\DorDsp\to f^+) \approx \Acp(\DorDsp\to f^+) + A_P(\DorDsp) + A_D(f^+),\label{eq:Araw_CS}\\
\end{equation}
where $A_P(\DorDsp)$ is the asymmetry of the \DorDsp-meson production cross-section~\cite{LHCb-PAPER-2012-026,LHCb-PAPER-2018-010} and $A_D(f^+)$ is the asymmetry of the reconstruction efficiency for the final state $f^+$. When $f^+=\KS h^+$ (with $h=K,\pi$), the detection asymmetry receives contributions from the $h^+$ hadron (indicated as companion hadron in the following), $A_D(h^+)$, and from the neutral kaon, $A_D(\Kzb)$. Relevant instrumental effects contributing to $A_D(h^+)$ may include differences in interaction cross-sections with matter between positive and negative hadrons and the slightly charge-asymmetric performance of the reconstruction algorithms. The contribution to $A_D(\Kzb)$ arises from \Kz and \Kzb mesons having different interaction cross-sections with matter and from their propagation in the detector being affected by the presence of \CP violation in the \Kz--\Kzb system. When $f^+=\phi(\to\Kp\Km)\pip$, the detection asymmetry is mostly due to the charged pion, as the contributions from the oppositely charged kaons cancel to a good precision.

The detection and production asymmetries are canceled by using the decays \mbox{\DKSpi}, \DsKSK and \Dsphipi, which proceed through the Cabibbo-favored $\cquark \to \squark\dquarkbar\uquark$ transition. In the SM, these decays are expected to have \CP asymmetries that are negligibly small compared to the Cabibbo-suppressed modes, when effects induced by the neutral kaons are excluded~\cite{Lipkin:1999qz,Yu:2017oky}. Hence, their raw asymmetries can be approximated as in Eq.~\eqref{eq:Araw_CS}, but with $\Acp=0$. The \CP asymmetries of the decay modes of interest are determined by combining the raw asymmetries as follows:
\begin{align}
\Acp(\DsKSpi) &\approx\ARAW(\DsKSpi)-\ARAW(\Dsphipi),\label{eq:ACP_DsKSpi}\\
\Acp(\DKSK)   &\approx\ARAW(\DKSK)-\ARAW(\DKSpi)\nonumber\\
              &\qquad-\ARAW(\DsKSK)+\ARAW(\Dsphipi),\label{eq:ACP_DKSK}\\
\Acp(\Dphipi) &\approx\ARAW(\Dphipi)-\ARAW(\DKSpi),\label{eq:ACP_Dphipi}
\end{align}
where the contribution from $A_D(\Kzb)$ is omitted and should be subtracted from any of the measured asymmetries where it is present.

The LHCb detector~\cite{Alves:2008zz,LHCb-DP-2014-002} is a single-arm forward spectrometer designed for the study of particles containing \bquark or \cquark quarks. The detector elements that are particularly relevant to this analysis are: a silicon-strip vertex detector that allows for a precise measurement of the impact parameter, \ie, the minimum distance of a charged-particle trajectory to a \proton\proton interaction point (primary vertex); a tracking system that provides a measurement of the momentum of charged particles; two ring-imaging Cherenkov detectors that are able to discriminate between different species of charged hadrons; and a calorimeter system that is used for the identification of photons, electrons and hadrons. The polarity of the magnetic field is periodically reversed during data-taking to mitigate the differences between reconstruction efficiencies of oppositely charged particles.

The online event selection is performed by a trigger, which consists of a hardware stage followed by a two-level software stage. In between the two software stages, an alignment and calibration of the detector is performed in near real-time and their results are used in the trigger~\cite{LHCb-PROC-2015-011}. Events with candidate \DorDsp decays are selected by the hardware trigger by imposing either that one or more \DorDsp decay products are associated with large transverse energy deposits in the calorimeter or that the accept decision is independent of the \DorDsp decay products (\ie, it is caused by other particles in the event). In the first level of the software trigger, one or more \DorDsp decay products must have large transverse momentum and be inconsistent with originating from any primary vertex. In the second level, the candidate decays are fully reconstructed using kinematic, topological and particle-identification criteria. The $\DorDsp\to\KS h^+$ candidates are made by combining charged hadrons with $\KS\to\pip\pim$ candidates that decay early enough for the final-state pions to be reconstructed in the vertex detector. This requirement suppresses to a negligible level possible \CP-violation effects due to interference between Cabibbo-favored and doubly Cabibbo-suppressed amplitudes with neutral-kaon mixing in the control-sample decays \DKSpi and \DsKSK~\cite{Yu:2017oky}.

The \DorDsp candidates reconstructed in the trigger are used directly in the offline analysis~\cite{LHCb-DP-2012-004,LHCb-DP-2016-001}. The candidates with a \KS meson in the final state are further selected offline using an artificial neural network~(NN), based on the multilayer perceptron algorithm~\cite{Hocker:2007ht}, to suppress background due to random combinations of \KS mesons and hadrons not originating from a $\DorDsp\to\KS h^+$ decay. The quantities used in the NN to discriminate signal from combinatorial background are: the \KS candidate momentum; the transverse momenta of the \DorDsp candidate and of the companion hadron; the angle between the \DorDsp candidate momentum and the vector connecting the primary and secondary vertices; the quality of the secondary vertex; and the track quality of the companion hadron. The NN is trained using signal and background data samples, obtained with the \sPlot method~\cite{Pivk:2004ty}, from a $\order(1\%)$ fraction of candidates randomly sampled. In the \DsKSpi case, thanks to similar kinematics, background-subtracted \DKSpi decays are exploited as a signal proxy to profit from larger yields. The thresholds on the NN response are optimized for the \DsKSpi and \DKSK decays by maximizing the value of $S/\sqrt{S+B}$, where $S$ and $B$ stands for the signal and background yield observed in the mass ranges $1.93<m(\KS\pi^+)<2.01$\gevcc and $1.83<m(\KS K^+)<1.91$\gevcc, respectively. Candidate $\DorDsp\to\phi(\to\Kp\Km)\pip$ decays are selected offline with requirements on the transverse momenta of the \DorDsp candidate and of the companion hadron, on the quality of the secondary vertex, and on the $\Kp\Km$ mass to be within $10$\mevcc of the nominal $\phi(1020)$-meson mass~\cite{PDG2018}. The mass window is chosen considering that the observed width is dominated by the $\phi(1020)$-meson natural width of $4.2\mevcc$~\cite{PDG2018} and is only marginally affected by the experimental resolution of $1.3\mevcc$.

\begin{figure}[hbt!]
\centering
\includegraphics[width=0.5\textwidth]{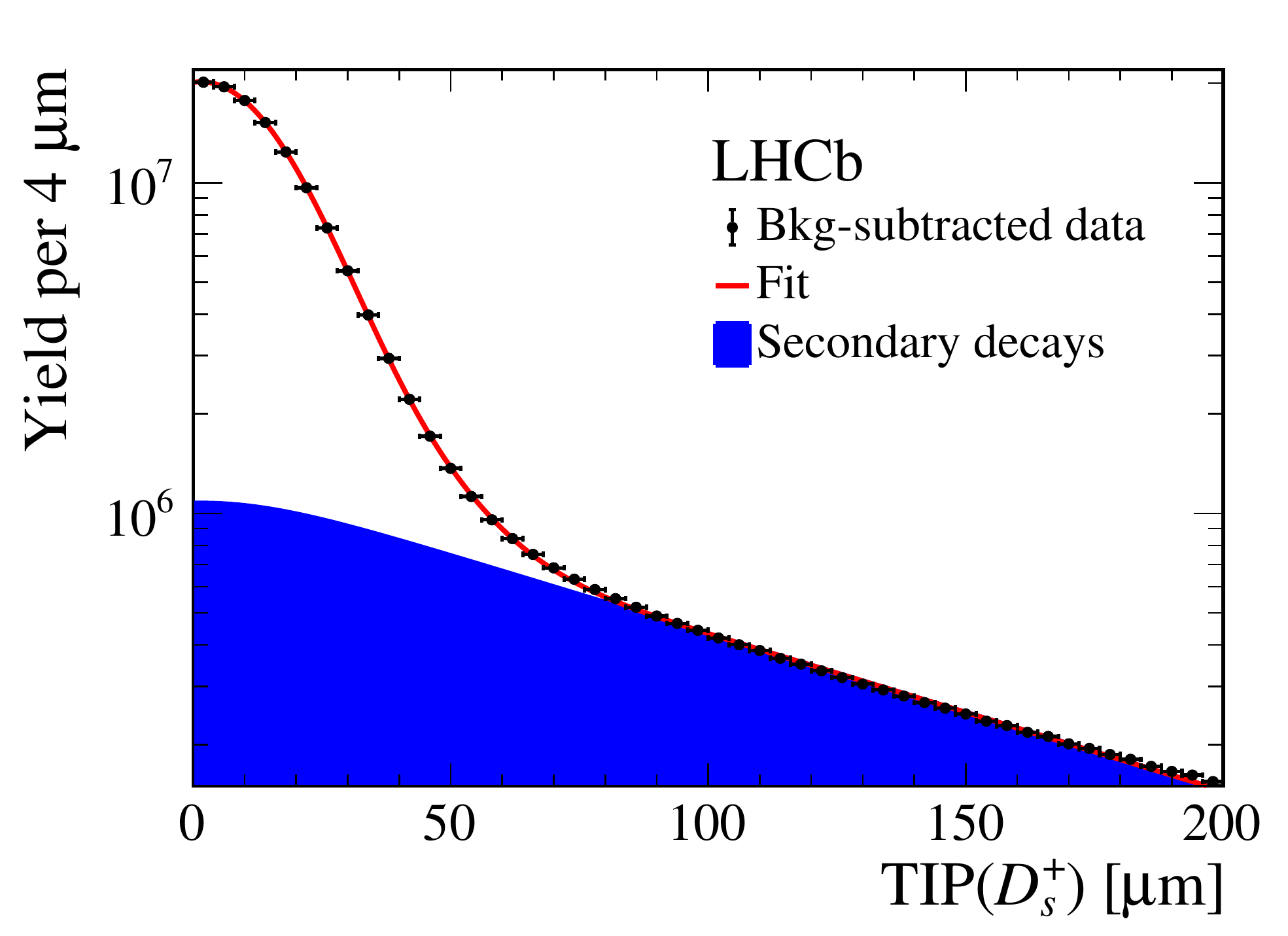}
\caption{Distribution of the transverse impact parameter (TIP) for background-subtracted \DsKSpi candidates with fit projections overlaid.}\label{fig:tip}
\end{figure}

The contribution of \DorDsp mesons produced through decays of \bquark hadrons, referred to as secondaries throughout, is suppressed by requiring that the \DorDsp impact parameter in the plane transverse to the beam (\tip) is smaller than $40\mum$. The remaining percent-level contribution is evaluated by means of a fit to the \tip distribution when such requirement is released, as shown in Fig.~\ref{fig:tip} for the \DsKSpi decay. The impact of the secondary background on the results is accounted for in the systematic uncertainties.

Typical sources of background from \DorDsp meson and \Lc baryon decays are: the \DsKSK and $\Lc\to\KS\proton$ decays, where the kaon and the proton are misidentified as a pion, when the signal is the \DsKSpi decay; the \DKSpi and $\Lc\to\KS\proton$ decays, where the pion and the proton are misidentified as a kaon, in the \DKSK case; and the $\Lc\to\phi\proton$ decay, where the proton is misidentified as a pion, when the signal is the \Dphipi decay. These are all reduced to a negligible level using particle-identification requirements and kinematic vetos.

Fiducial requirements are imposed to exclude kinematic regions that induce a large asymmetry in the companion-hadron reconstruction efficiency. These regions occur because low momentum particles of one charge at large (small) angles in the bending plane may be deflected out of the detector acceptance (into the noninstrumented beam pipe region), whereas particles with the other charge are more likely to remain within the acceptance. About 78\%, 93\% and 94\% of the selected candidates are retained by these fiducial requirements for \DorDspKSpi, \DorDspKSK and \DorDspphipi decays, respectively.  

Detection and production asymmetries may depend on the kinematics of the involved particles. Therefore, the cancellation provided by the control decays is accurate only if the kinematic distributions agree between any pair of signal and control modes, or pair of control modes entering Eqs.~\eqref{eq:ACP_DsKSpi}--\eqref{eq:ACP_Dphipi}. Differences are observed, and the ratio between background-subtracted~\cite{Pivk:2004ty} signal and control sample distributions of transverse momentum, azimuthal angle and pseudorapidity are used to define candidate-by-candidate weights. The background-subtracted candidates of the control decays are weighted such that their distributions agree with those of the signal using an iterative procedure. The process consists of calculating the weights in each one-dimensional distribution of the weighting variables and repeating the procedure until good agreement is achieved among all the distributions. For the measurements of the \DsKSpi and \Dphipi\ \CP asymmetries, the \Dsphipi and \DKSpi control samples are weighted so that the \DorDsp meson and companion-pion kinematic distributions agree with their respective signal samples to cancel the \DorDsp production and companion-pion detection asymmetries. In the case of the $\Acp(\DKSK)$ measurement, the \Dp kinematic distributions of the \DKSpi sample are weighted to those of the \DKSK signal to cancel the \Dp production asymmetry, and the $\Kp$ distributions of the \DsKSK decays are weighted to those of the \DKSK signal to cancel the kaon detection asymmetry. The \DKSpi and \DsKSK control decays then introduce their own additional nuisance asymmetries, which need to be corrected for using the \Dsphipi control decay. Hence, the \Dsp and companion-pion kinematic distributions of the \Dsphipi sample are made to agree with those of the \DsKSK and \DKSpi samples, respectively, to cancel the \Dsp production and companion-pion detection asymmetries.

\begin{figure}[t]
\centering
\begin{overpic}[width=0.5\textwidth]{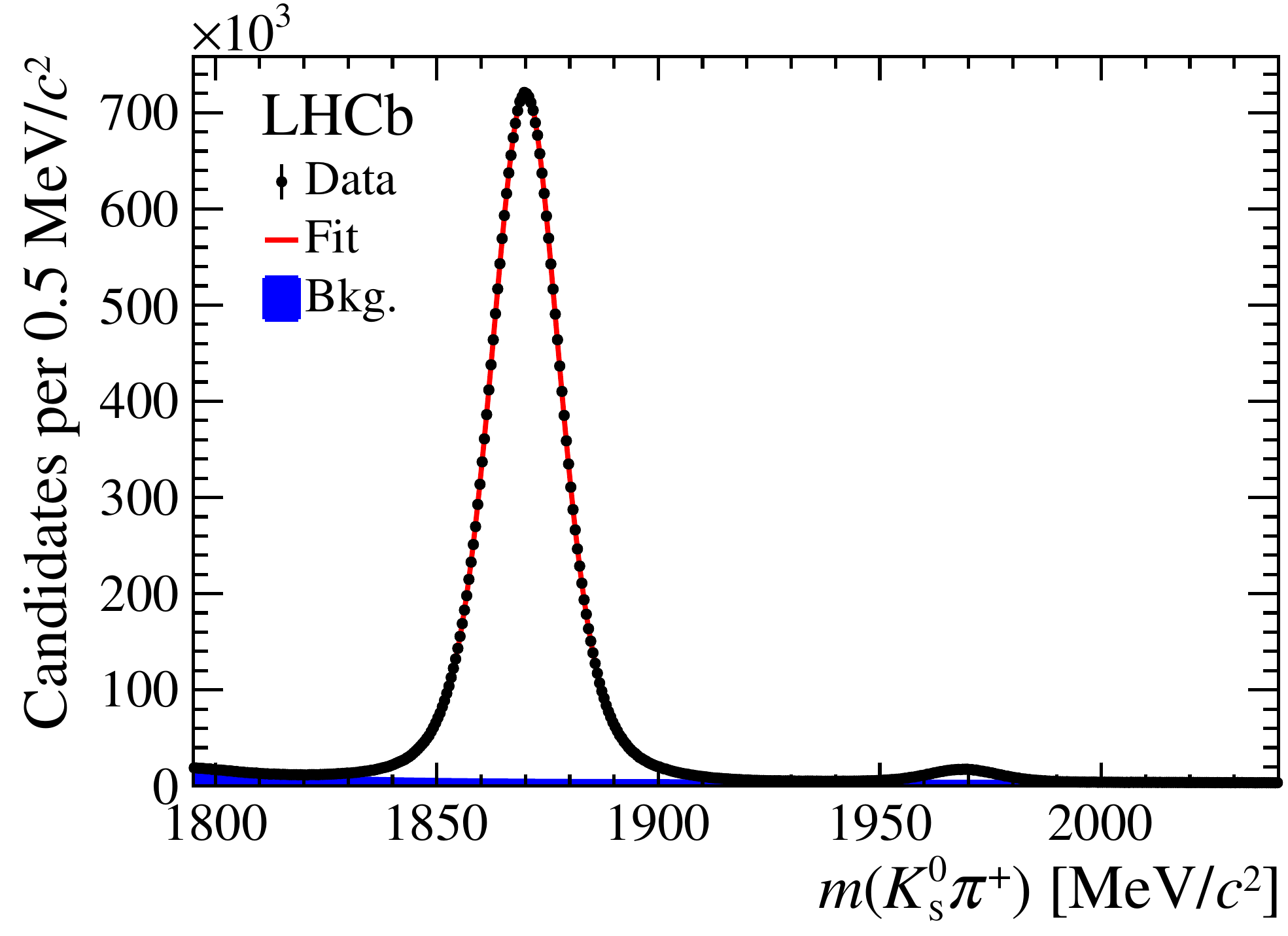}
\put(50,22){\includegraphics[scale=0.18]{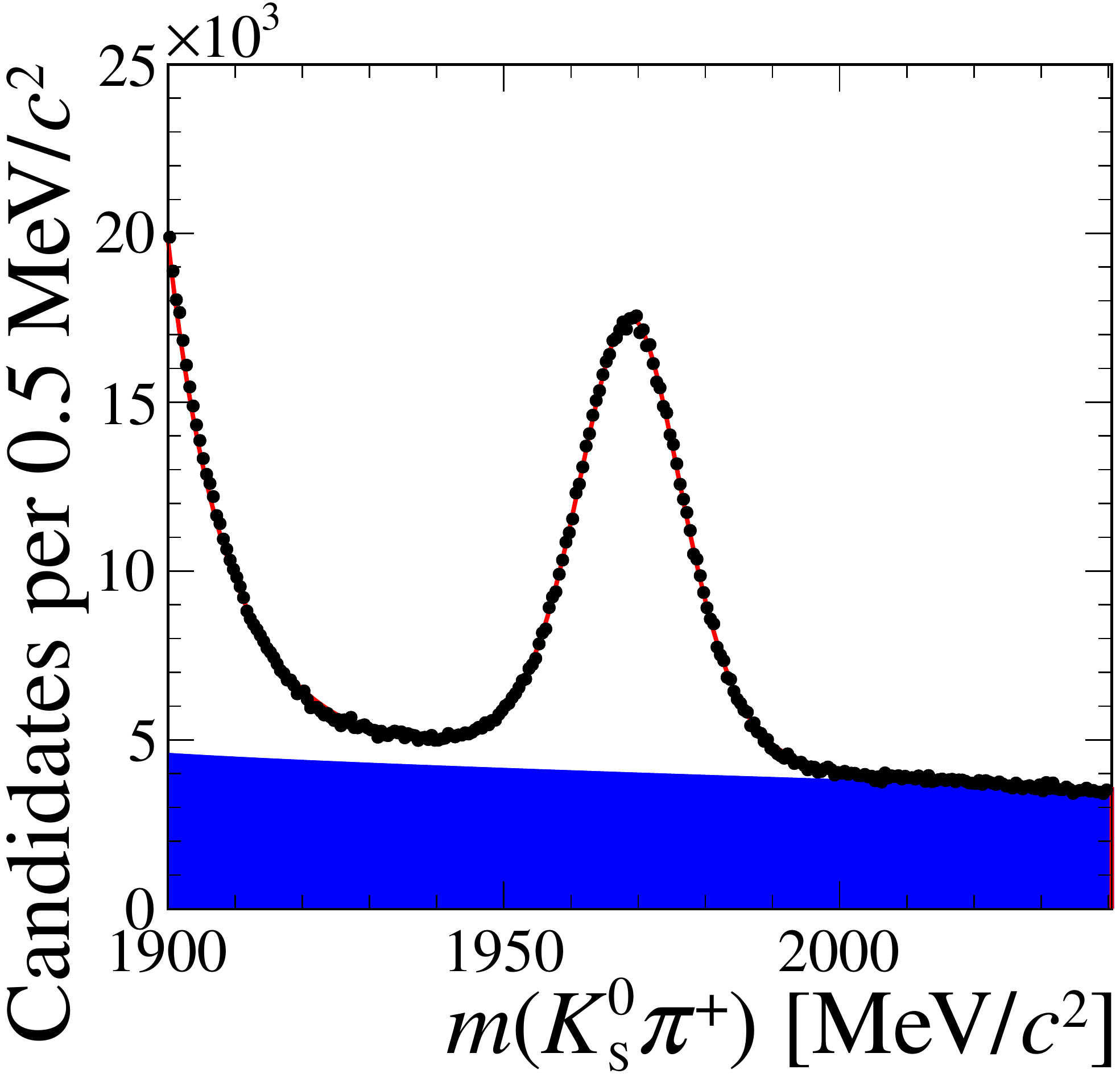}}
\end{overpic}\\
\includegraphics[width=0.5\textwidth]{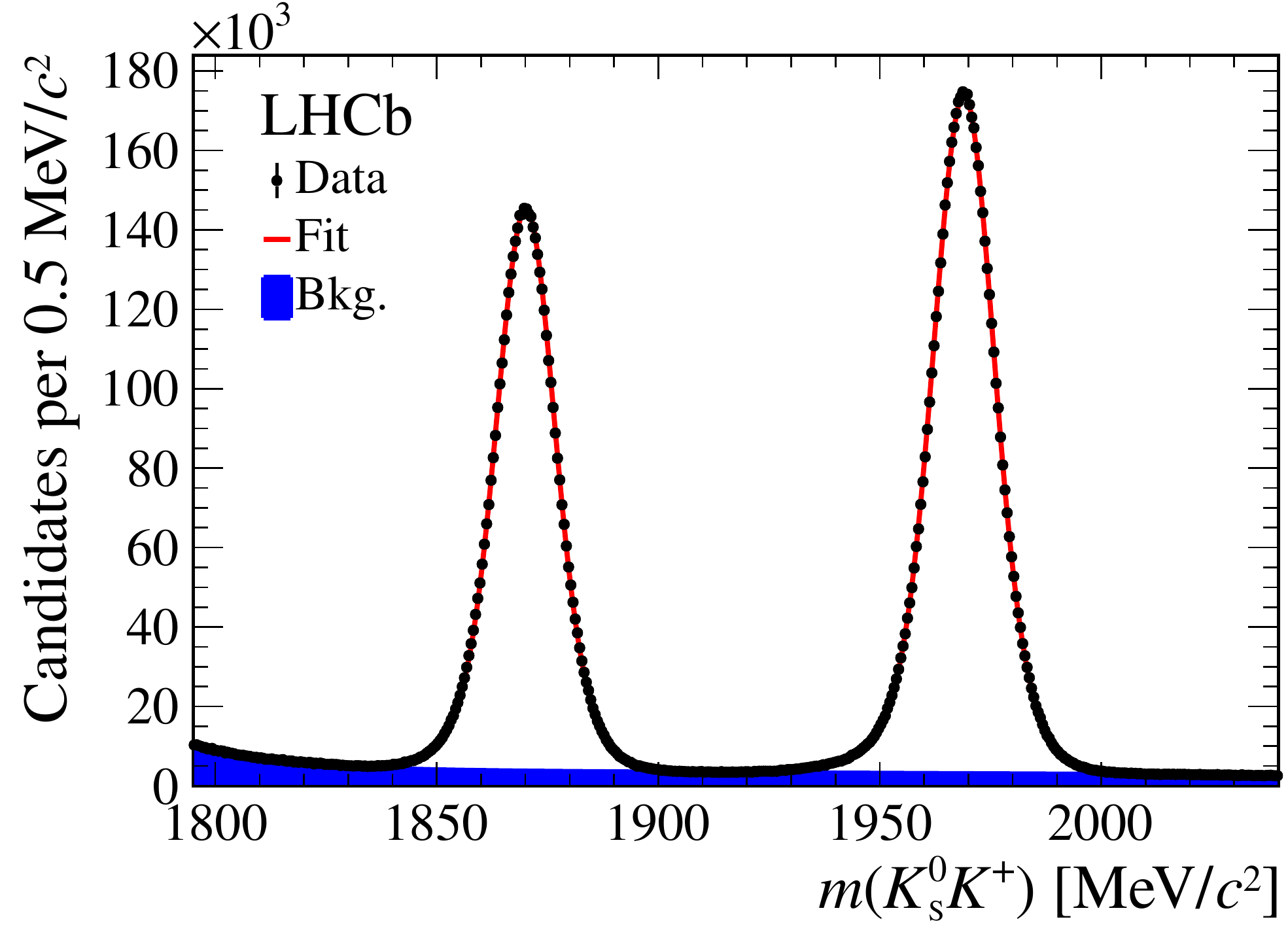}\\
\includegraphics[width=0.5\textwidth]{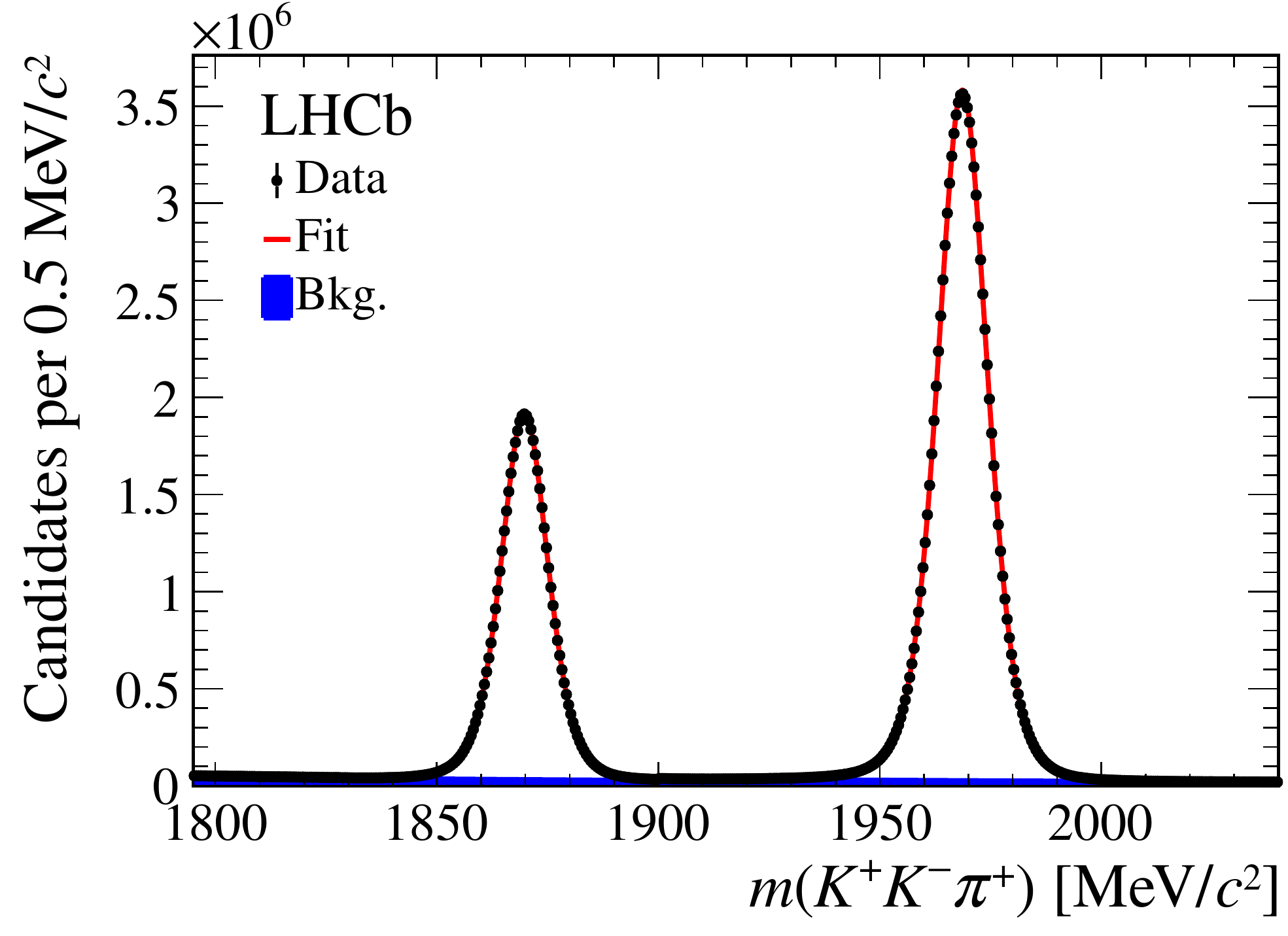}\\
\caption{Mass distributions of the selected (top) $\DorDsp\to\KS\pip$, (middle) $\DorDsp\to\KS\Kp$ and (bottom) $\DorDsp\to\phi\pip$ candidates with fit projections overlaid. The inset in the top plot shows the mass distribution around the \DsKSpi signal region.\label{fig:dm}}
\end{figure}

Simultaneous least-squares fits to the mass distributions of weighted \DorDsp and \DorDsm candidates determine the raw asymmetries for each decay mode considered. To avoid experimenter bias, the raw asymmetries of the Cabibbo-suppressed signals were shifted by unknown offsets sampled uniformly between $-1\%$ and $1\%$, such that the results remained blind until the analysis procedure was finalized. In the fits, the signal and control decays are modeled as the sum of a Gaussian function to describe the core of the peaks, and a Johnson $S_U$ distribution~\cite{Johnson:1949zj}, which accounts for the asymmetric tails. The combinatorial background is described by the sum of two exponential functions. All shape parameters are determined from the data. In each fit, signal and control decays share the same shape parameters apart from a mass shift, which accounts for the known difference between the \Dsp and \Dp masses~\cite{PDG2018}, and a relative scale factor between the peak widths, which is also determined from the data. The means and widths of the peaks, as well as all background shape parameters, are allowed to differ between \DorDsp and \DorDsm decays. The projections of the fits to the combined \DorDsp and \DorDsm data are shown in Fig.~\ref{fig:dm}. The samples contain approximately $600$ thousand \DsKSpi, $5.1$ million \DKSK, and $53.3$ million \Dphipi signal candidates, together with approximately $30.5$ million \DKSpi, $6.5$ million \DsKSK, and $107$ million \Dsphipi control decays.

The raw asymmetries are, where relevant, corrected for the neutral-kaon detection asymmetry. The net correction is estimated following Ref.~\cite{LHCb-PAPER-2014-013} to be $(+0.084 \pm 0.005)\%$ for $\Acp(\DsKSpi)$, $(-0.086 \pm 0.005)\%$ for $\Acp(\DKSK)$, and $(-0.068 \pm 0.004)\%$ for $\Acp(\Dphipi)$, where the uncertainty is dominated by the accuracy of the detector modeling in the simulation. The asymmetries are combined following Eqs.~\eqref{eq:ACP_DsKSpi}--\eqref{eq:ACP_Dphipi} to obtain $\Acp(\DsKSpi)=\left(\AcpDsKSpiRes\pm\AcpDsKSpiStat\right)\times\AcpDsKSpiUnit$, 
$\Acp(\DKSK) =\left(\AcpDKSKRes\pm\AcpDKSKStat\right)\times\AcpDKSKUnit$,  $\Acp(\Dphipi)=\left(\AcpDphipiRes\pm\AcpDphipiStat\right)\times\AcpDphipiUnit$, where the uncertainties are only statistical.

Several sources of systematic uncertainty affecting the measurement are considered as reported in Table~\ref{tab:syst}. The dominant contribution is due to the assumed shapes in the mass fits. This is evaluated by fitting with the default model large sets of pseudoexperiments where alternative models that describe data equally well are used in generation. For $\Acp(\DsKSpi)$ and $\Acp(\DKSK)$, the second leading contribution is due to the residual contamination from secondary \DorDsp decays, which introduces a small difference between the asymmetry of \DorDsp-meson production cross-sections of the signal and control modes. For \mbox{$\Acp(\Dphipi)$}, instead, the second leading systematic uncertainty arises from neglected kinematic differences between the $\phi$-meson decay products. These differences, mainly caused by the interference between the $S$-wave and $\phi\pip$ decay amplitudes in the $\Kp\Km$-mass region under study, result in an imperfect cancelation of the charged-kaon detection asymmetry. Other subleading contributions are due to the inaccuracy in the equalization of the kinematic distributions between signal and control samples, and to the uncertainty in the neutral-kaon detection asymmetry. 

In addition, several consistency checks are performed to investigate possible unexpected biases by comparing results obtained in subsamples of the data defined according to the data-taking year and magnetic-field polarity, the per-event track multiplicity, the configurations of the hardware- and software-level triggers, and the \DorDsp momentum. A \chisq test has been performed for each cross-check and the corresponding $p$ values are consistent with being uniformly distributed; the lowest (largest) $p$ value is $4\%$ ($86\%$). Therefore, the observed variations in results are consistent with statistical fluctuations and no additional sources of systematic uncertainties are considered.

\begin{table}[t]
\centering
\caption{Summary of the systematic uncertainties (in units of $10^{-3}$) on the measured quantities. The total is the sum in quadrature of the different contributions.\label{tab:syst}}
\resizebox{\textwidth}{!}{
\begin{tabular}{lccc}
\toprule
Source & $\Acp(\DsKSpi)$ & $\Acp(\DKSK)$ & $\Acp(\Dphipi)$ \\
\midrule
Fit model          & 0.39 & 0.44 & 0.24 \\
Secondary decays   & 0.30 & 0.12 & 0.03 \\
Kinematic differences  & 0.09 & 0.09 & 0.04 \\
Neutral kaon asymmetry & 0.05 & 0.05 & 0.04 \\
Charged kaon asymmetry & 0.08 & 0.09 & 0.15 \\
\midrule
Total              & 0.51 & 0.48 & 0.29 \\
\bottomrule
\end{tabular}}
\end{table}

In summary, using proton-proton collision data collected with the LHCb detector at a center-of-mass energy of 13\tev, and corresponding to $3.8\invfb$ of integrated luminosity, the following \CP asymmetries are measured:
\begin{align*}
\Acp(\DsKSpi)&=\left(\phantom{-}\AcpDsKSpiRes\phantom{0}\pm\AcpDsKSpiStat\phantom{0}\pm\AcpDsKSpiSyst\phantom{0}\right)\times\AcpDsKSpiUnit,\\
\Acp(\DKSK)  &=\left(\AcpDKSKRes\pm\AcpDKSKStat\pm\AcpDKSKSyst\right)\times\AcpDKSKUnit,\\
\Acp(\Dphipi)&=\left(\phantom{-}\AcpDphipiRes\pm\AcpDphipiStat\pm\AcpDphipiSyst\right)\times\AcpDphipiUnit,
\end{align*}
where the first uncertainties are statistical and the second systematic. Effects induced by \CP violation in the neutral kaon system are subtracted from the measured asymmetries. The results represent the most precise determination of these quantities to date and are consistent with \CP symmetry. They are in agreement with previous LHCb determinations based on independent data samples collected at center-of-mass energies of 7 and 8\tev~\cite{LHCb-PAPER-2012-052,LHCb-PAPER-2014-018}, as well as with measurements from other experiments~\cite{Mendez:2009aa,Ko:2010ng,Lees:2012jv,Ko:2012uh,Lees:2012nn,Staric:2011en}. The results are combined with previous LHCb measurements using the BLUE method~\cite{Lyons:1988rp}. The systematic uncertainties are considered uncorrelated, apart from those due to the neutral- and charged-kaon detection asymmetries that are fully correlated. The combination yields
\begin{align*}
\Acp(\DsKSpi)&=\left(\phantom{-}1.6\phantom{0}\pm 1.7\phantom{0}\pm 0.5\phantom{0}\right)\times\AcpDsKSpiUnit,\\
\Acp(\DKSK)&=\left(-0.04\pm 0.61\pm 0.45\right)\times\AcpDKSKUnit,\\
\Acp(\Dphipi)&=\left(\phantom{-}0.03\pm 0.40\pm 0.29\right)\times\AcpDphipiUnit,
\end{align*}
where the first uncertainties are statistical and the second systematic. No evidence for \CP violation in these decays is found. More precise measurements of these asymmetries can be expected when the data already collected by LHCb in 2018 are included in a future analysis, and when much larger samples will become available at the upgraded LHCb detector~\cite{LHCb-PII-Physics}.

\section*{Acknowledgements}
%
%
\noindent We express our gratitude to our colleagues in the CERN
accelerator departments for the excellent performance of the LHC. We
thank the technical and administrative staff at the LHCb
institutes.
We acknowledge support from CERN and from the national agencies:
CAPES, CNPq, FAPERJ and FINEP (Brazil); 
MOST and NSFC (China); 
CNRS/IN2P3 (France); 
BMBF, DFG and MPG (Germany); 
INFN (Italy); 
NWO (Netherlands); 
MNiSW and NCN (Poland); 
MEN/IFA (Romania); 
MSHE (Russia); 
MinECo (Spain); 
SNSF and SER (Switzerland); 
NASU (Ukraine); 
STFC (United Kingdom); 
NSF (USA).
We acknowledge the computing resources that are provided by CERN, IN2P3
(France), KIT and DESY (Germany), INFN (Italy), SURF (Netherlands),
PIC (Spain), GridPP (United Kingdom), RRCKI and Yandex
LLC (Russia), CSCS (Switzerland), IFIN-HH (Romania), CBPF (Brazil),
PL-GRID (Poland) and OSC (USA).
We are indebted to the communities behind the multiple open-source
software packages on which we depend.
Individual groups or members have received support from
AvH Foundation (Germany);
EPLANET, Marie Sk\l{}odowska-Curie Actions and ERC (European Union);
ANR, Labex P2IO and OCEVU, and R\'{e}gion Auvergne-Rh\^{o}ne-Alpes (France);
Key Research Program of Frontier Sciences of CAS, CAS PIFI, and the Thousand Talents Program (China);
RFBR, RSF and Yandex LLC (Russia);
GVA, XuntaGal and GENCAT (Spain);
the Royal Society
and the Leverhulme Trust (United Kingdom);
Laboratory Directed Research and Development program of LANL (USA).

\addcontentsline{toc}{section}{References}
\bibliographystyle{LHCb}
\bibliography{main,standard,LHCb-PAPER,LHCb-CONF,LHCb-DP,LHCb-TDR}

 
\clearpage
\centerline{\large\bf LHCb collaboration}
\begin{flushleft}
\small
R.~Aaij$^{28}$,
C.~Abell{\'a}n~Beteta$^{46}$,
B.~Adeva$^{43}$,
M.~Adinolfi$^{50}$,
C.A.~Aidala$^{77}$,
Z.~Ajaltouni$^{6}$,
S.~Akar$^{61}$,
P.~Albicocco$^{19}$,
J.~Albrecht$^{11}$,
F.~Alessio$^{44}$
% ...

{\footnotesize \it
% ...
$ ^{6}$Universit{\'e} Clermont Auvergne, CNRS/IN2P3, LPC, Clermont-Ferrand, France\\
% ...
$ ^{11}$Fakult{\"a}t Physik, Technische Universit{\"a}t Dortmund, Dortmund, Germany\\
% ...
$ ^{19}$INFN Laboratori Nazionali di Frascati, Frascati, Italy\\
% ...
$ ^{28}$Nikhef National Institute for Subatomic Physics, Amsterdam, Netherlands\\
% ...
$ ^{43}$Instituto Galego de F{\'\i}sica de Altas Enerx{\'\i}as (IGFAE), Universidade de Santiago de Compostela, Santiago de Compostela, Spain\\
$ ^{44}$European Organization for Nuclear Research (CERN), Geneva, Switzerland\\
% ...
$ ^{46}$Physik-Institut, Universit{\"a}t Z{\"u}rich, Z{\"u}rich, Switzerland\\
% ...
$ ^{50}$H.H. Wills Physics Laboratory, University of Bristol, Bristol, United Kingdom\\
% ...
$ ^{61}$University of Cincinnati, Cincinnati, OH, United States\\
% ...
$ ^{63}$Syracuse University, Syracuse, NY, United States\\
% ...
$ ^{77}$University of Michigan, Ann Arbor, United States, associated to $^{63}$\\
% ...
}
\end{flushleft}

\end{document}